# Stress Induced Structural Transformations in Au Nanocrystals

Abhinav Parakh[1], Sangryun Lee[2], Mehrdad T. Kiani[1], David Doan[3], Martin Kunz[4], Andrew Doran[4], Seunghwa Ryu[2] and X. Wendy Gu[3]*

[1]Materials Science and Engineering, Stanford University, Stanford, CA 94305, USA.
[2]Mechanical Engineering, KAIST, Yuseong-gu, Daejeon 34141, Republic of Korea.
[3]Mechanical Engineering, Stanford University, Stanford, CA 94305, USA.
[4]Advanced Light Source, Lawrence Berkeley National Lab, Berkeley 94720, USA.

**Abstract:** Nanocrystals can exist in multiply twinned structures like the icosahedron, or single crystalline structures like the cuboctahedron or Wulff-polyhedron. Structural transformation between these polymorphic structures can proceed through diffusion or displacive motion. Experimental studies on nanocrystal structural transformations have focused on high temperature diffusion mediated processes. Thus, there is limited experimental evidence of displacive motion mediated structural transformations. Here, we report the high-pressure structural transformation of 6 nm Au nanocrystals under nonhydrostatic pressure in a diamond anvil cell that is driven by displacive motion. *In-situ* X-ray diffraction and transmission electron microscopy were used to detect the transformation of multiply twinned nanocrystals into single crystalline nanocrystals. High-pressure single crystalline nanocrystals were recovered after unloading, however, the nanocrystals quickly reverted back to multiply twinned state after redispersion in toluene solvent. The dynamics of recovery was captured using transmission electron microscopy which showed that the recovery was governed by surface recrystallization and rapid twin boundary motion. We show that this transformation is energetically favorable by calculating the pressure-induced change in strain energy. Molecular dynamics simulations showed that defects nucleated from a region of high stress region in the interior of the nanocrystal, which make twin boundaries unstable. Deviatoric stress driven Mackay transformation and dislocation/disclination mediated detwinning are hypothesized as possible mechanisms of high-pressure structural transformation.

Metallic nanocrystals are used widely in fields such as photonics, biomedical therapies, catalysis, electronics and sensing[1]. Properties of these nanocrystals are highly dependent on their size, shape, and crystalline structure[2]. Multiply twinned (MT) icosahedron, MT decahedron, single-crystal (SC) cuboctahedron, and SC Wulff-polyhedron nanocrystal shapes are commonly observed, and can have different catalytic, magnetic, mechanical, structural, and electronic properties[3–8]. For this reason, it is often desirable to synthesize one particular nanocrystal size and shape, and maintain this structure during use. This remains difficult because the thermodynamic stability and structural transitions between different nanocrystal structures are still incompletely understood. The structural transformation of polyhedral structures such as MT icosahedron is also important for understanding materials like metallic glasses and magnetic nanoclusters, in which polyhedral atomic clusters make up the basic structural unit, and changes in these atomic clusters dictate material properties[9–11].

Structural transformation between different nanocrystal shapes have been studied using theory, simulations, and experiments. Using energy balance calculations and molecular dynamics (MD) simulations that consider differences in surface energy and lattice strain, it has been determined that MT nanocrystals are stable at smaller sizes and SC nanocrystals are stable at larger sizes[9,12–14]. The transition occurs from 2 to 10 nm depending on the calculation method, and varies in experiments due to the influence of surface ligands, solvents and substrates on surface energy. It has been proposed that the transformation between MT and SC structures occurs through diffusive or displacive processes, such as surface melting and restructuring, dislocation/disclination activity, and the Mackay transformation[15–18]. Transformation in nanocrystals have been studied experimentally by heating nanocrystals with the electron beam in a transmission electron microscope (TEM), high energy laser pulses, and annealing nanocrystals on a substrate[15,16,19–22]. These experimental studies observed that enhanced mobility, melting and recrystallization of nanocrystals lead to diffusion mediated structural transformations. However, displacive motion mediated structural transformation has not been studied systematically in nanocrystals.

High-pressure compression in a diamond anvil cell (DAC) is an ideal technique to study displacive motion in nanomaterials, because diffusion is suppressed at high pressure[23]. DAC has previously been used to study high-pressure phase transformation, crystallization and sintering of aggregated nanocrystals[24]. DAC techniques have also been used to study structural transformations in Ag nanocrystals under hydrostatic pressures,[25] which minimizes both diffusion and displacive motion. Here, we study the structural stability and structural transformation between MT and SC nanocrystals by compressing 6 nm Au nanocrystals in a DAC under non-hydrostatic pressure, and monitoring nanocrystal structure using *in-situ* X-ray diffraction (XRD). The nanocrystals are recovered after compression and imaged using TEM. We find that the 6 nm nanocrystals undergo a MT to SC transformation after compression to 7.7 GPa of pressure. This is in contrast to smaller, 3.9 nm Au nanocrystals which did not show a structural transformation under pressure, and instead formed stacking faults via surface nucleated partial dislocations[26]. MD simulations were conducted to understand defect formation in nanocrystals of 3.9 nm and 6 nm in size. These simulations showed that dislocation activity is enhanced in larger nanocrystals. These results indicate that displacive motion driven large scale structural transformation is possible in nanocrystals and must be considered in designing structures at the nanoscale.

MT Au nanocrystals were synthesized using organic phase reduction of chloroauric acid and capped with dodecanethiol ligands[27]. The nanocrystal size distribution was found to be 6.0±0.3 nm using TEM (see Fig. 1A and Fig. S1). High-resolution TEM images showed that the majority of nanocrystals (~80%) were MT and remaining nanocrystals were SC (a total of 59 nanocrystals were analyzed). The MT nanocrystals were icosahedral structures which are formed with 20 tetrahedral units joined by 20 twin boundaries. An icosahedral polyhedron has 6 5-fold, 10 3-fold, and 15 2-fold axes. Fig. 1B shows the icosahedral nanocrystal along the 3-fold axis and Fig. 1C shows the icosahedral nanocrystal along a 2-fold axis. The SC nanocrystals were cuboctahedron or Wulff-polyhedron in structure, and sometimes contained 1-2 twin boundaries rather than the high density of twin boundaries in MT nanocrystals.

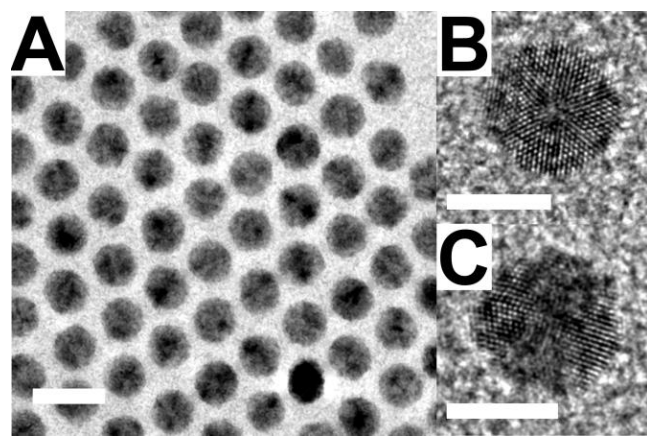

Fig. 1. TEM images of 6 nm Au nanocrystals. A) Bright field image of monodisperse nanocrystals. Scale bar is 10 nm. B, C) High-resolution images of icosahedral nanocrystals. Scale bar is 5 nm.

Ambient pressure XRD for the nanocrystals showed an FCC crystal structure, and significantly broader peaks than bulk Au due to crystallite size broadening (see Fig. S2). Nanocrystal surfaces exert a Laplace pressure on the interior of the nanocrystal, which scales inversely with the radius[28]. This compressive force shifts all the ambient pressure XRD peaks except the (200) peak to a higher 2θ angle compared to the bulk. The {111} planes form the surface of MT icosahedral nanocrystals. Hence, the (111) peak was shifted by ~0.06° 2θ compared to the bulk, which corresponds to a volumetric strain of ~1.5%. The position of the (200) peak does not shift in the same way as the other peaks because it is affected by twinning in the nanocrystal. This was previously shown in a model which revealed that the (200) peak shifts towards lower 2θ angles with an increase in twinning density[29,30]. This model simulates the effect of low twinning density and cannot be directly applied to MT nanocrystals which each contain 20 twins, but the qualitative trend is still relevant. Another feature of the (200) peak is the double peak which is due to the mixture of 80% MT and 20% SC nanocrystals. One peak is located at the bulk (200) peak position, and the other is shifted towards lower 2θ angles by ~0.6° 2θ. The icosahedral nanocrystals correspond to the lower 2θ (200) peak, which is shifted due to the twins, and the SC nanocrystals correspond to the (200) peak at the bulk position.

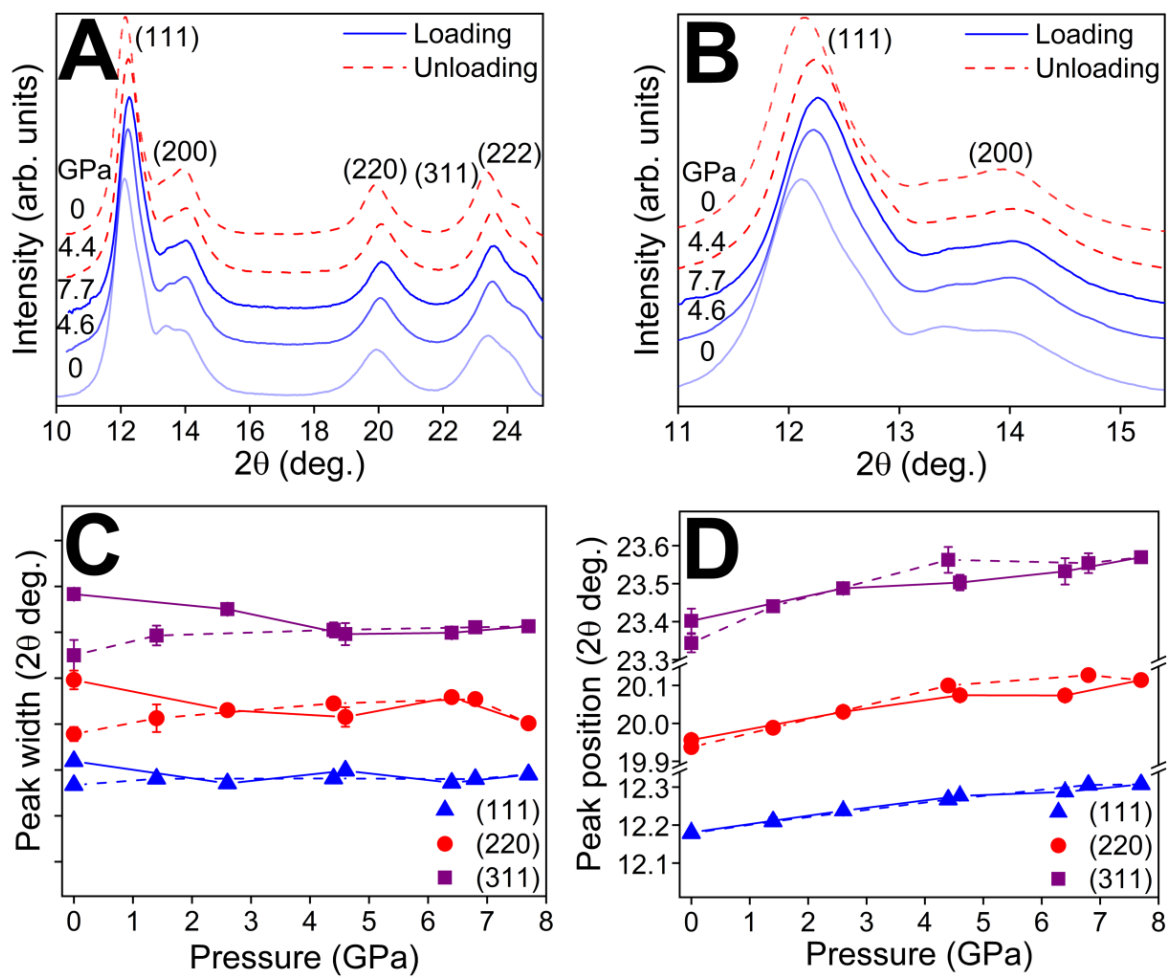

Fig. 2. High-pressure XRD for 6 nm nanocrystals. A) All diffraction peaks and B) magnified view of (111) and (200) peaks. Change in diffraction peak C) position and D) width (each division is 0.1°), upon loading (solid line) and unloading (dashed line).

High-pressure XRD was obtained *in-situ* during DAC compression experiments at the Advanced Light Source at Lawrence Berkeley National Laboratory. Toluene was used as a non-hydrostatic pressure medium[31] and nanocrystals were loaded as a thick film at the bottom of the DAC sample chamber. XRD was collected while the nanocrystals were loaded up to 7.7 GPa and as pressure was released. The pressure was limited to 7.7 GPa to avoid sintering between the nanocrystals, which has been observed at higher pressures[32–34]. The XRD peak position and width (full width at half maximum) were observed to change with increasing and decreasing pressure and were quantified at each pressure (Fig. 2).

High-pressure XRD and the corresponding peak positions and widths are shown in Fig. 2. The shift in XRD peak position indicates the pressure-induced elastic strain in the nanocrystals. XRD peak position for all peaks except the (200) peak recovered completely with pressure cycling to within 0.1% of their original value (Fig. 2 D). An irreversible change was observed for the (200) peak position with pressure cycling (Fig. 2 B). The ratio of the left to the right (200) peak intensities is proportional to the degree of twinning, or the fraction of MT to SC nanocrystals in the sample[29]. After pressure cycling, this ratio decreased by ~22%: the right (200) peak intensity increased significantly with pressure and remained at higher values after unloading, while the left (200) peak decreased in intensity. This indicated that the MT nanocrystals detwinned with pressure cycling and underwent a structural transformation from MT to SC. Changes in peak width with pressure cycling also indicate that this structural transformation occurred (see Fig. 2 C). The XRD peak width for (111), (220) and (311) peaks decreased by 11%, 19%, and 22%, respectively. This can be explained by an increase in crystallite size upon transformation from MT to SC nanocrystals[35].

Post-compression TEM imaging corroborated these findings. Nanocrystals were loaded to ~5 GPa in the DAC. The sample was then quickly unloaded, and the sample chamber was opened to air to dry out the liquid toluene. The nanocrystals were picked up using a needle and scraped onto a TEM grid and inserted into the TEM within 10 minutes. The post-compression TEM images are shown in Fig. 3. We found that the ratio of nanocrystals changed from 80%

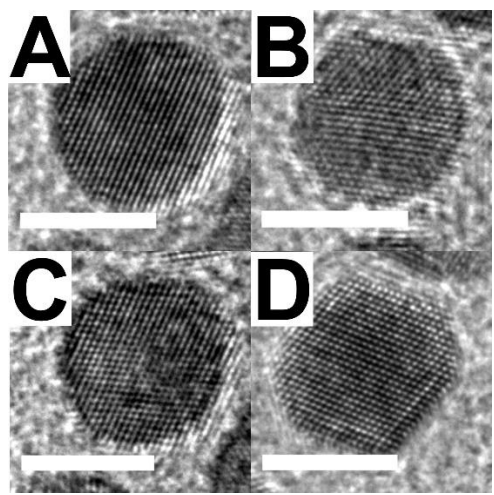

Fig. 3. Post-compression TEM images of transformed single crystalline 6 nm nanocrystals.

MT and 20% SC nanocrystals before pressure cycling, to 40% MT and 60% SC nanocrystals after pressure cycling. The fraction of MT nanocrystals decreased by 50% with pressure cycling. High-resolution TEM images of 59 as-synthesized and 23 post-compression nanocrystals were analyzed. Post-compression nanocrystals were SC with cuboctahedron, truncated-octahedron or Wulff-polyhedron shapes (Fig. 3). Some SC nanocrystals had a twin that extended across the nanocrystal (Fig. 3 B). Using the ratio of MT to SC nanocrystals from TEM, the Debye scattering equation was used to simulate pre- and post-compression XRD patterns. Fig. S4 shows the simulated XRD pattern for mixtures of 80:20 and 40:60 MT and SC nanocrystals. The simulated XRD pattern showed similar trends as the experimental XRD patterns, in which the ratio of the left and right (200) peaks decreased with decreasing fraction of MT nanocrystals. This showed that the post-compression TEM analysis matches the high-pressure XRD patterns.

The post-compression SC structure of the nanocrystal was observed to be unstable. Toluene was added drop by drop to a TEM grid with post-compression nanocrystals. TEM imaging was performed after waiting for 10-15 mins, which showed that the ratio of MT to SC structures reverted close to the as-synthesized value (85% MT and 15% SC, 48 nanocrystals analyzed). This showed that the nanocrystal can rapidly convert to the thermodynamically stable MT structure in solution at ambient pressure (see Fig. S5). The dynamics and mobility of twin boundaries in nanocrystals was further investigated by heating individual nanocrystals under a 200 keV electron beam within the TEM. TEM movie and snapshots of nanocrystals transforming are shown in Supplementary Movie S1 and Fig. 4. At the start of the movie, nanocrystal I is 7 nm in size and has two inclined twin boundaries at 35°. Nanocrystal II is 6.3 nm in size and half of nanocrystal II has no twin boundaries and the other half of it has a MT structure (Fig. 4 A). Fig. 4 B, C and D show the nanocrystals after 10 s, 40 s and 70 s of electron beam irradiation, respectively. After 10 s, nanocrystal I rapidly developed a MT structure in the lower half of the nanocrystal, and the angle between the twin boundaries increased to ~70°. The surface of nanocrystal I started melting and sintering with the nanocrystal II. After 40 s, the surface of nanocrystal II started melting and nanocrystal II rotated to sinter with the nanocrystal I. The twin boundaries in nanocrystal I dynamically moved away from the sintered part of the nanocrystal. Fig. 4 D shows final state of the nanocrystals. A SC region connects both nanocrystals. The nanocrystal I has a partial MT structure with the twin boundaries at an angle of ~71° which is close to the ideal ~72° for a strained penta-twinned structure. This showed that the twin boundaries in nanocrystal can evolve due to enhanced diffusion under excitation by the electron beam. It is likely that the enhanced mobility of twin boundaries and interaction of ligands/surface of the nanocrystal with toluene solvent resulted in the rapid recovery of MT structure from SC nanocrystal in solution. The post-compression TEM and high-pressure XRD analysis confirmed that the MT 6 nm nanocrystals transformed into SC nanocrystals with pressure cycling, and the SC structure was unstable at ambient pressure and reverted back to MT structure after leaving in solution for short time.

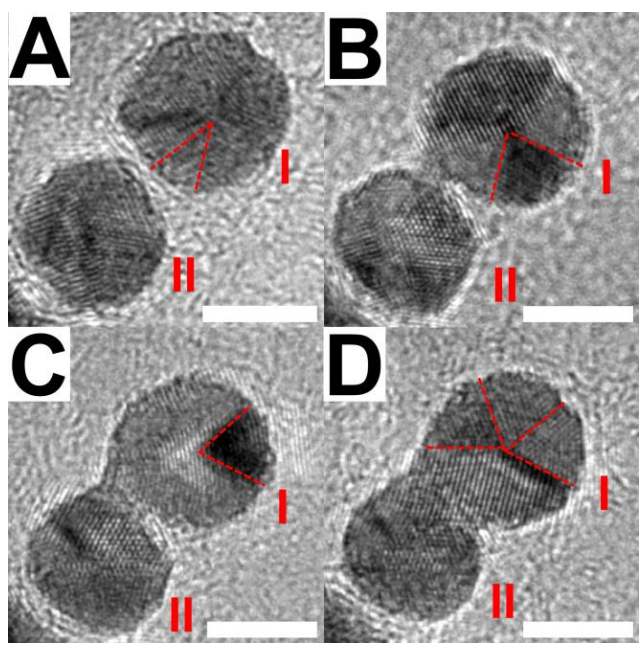

**Fig. 4. Snapshots from *in-situ* TEM movie showing transformation from SC back to MT under electron irradiation.** A) Nanocrystal I and II at the beginning of imaging and after B) 10 s, C) 40 s and D) 70 s of electron irradiation. Red dashed line denotes the twin boundary in nanocrystal I. Scale bar is 5 nm.

The high-pressure behavior of 6 nm nanocrystals differs from that of 3.9 nm nanocrystals previously studied by our group[26]. High pressure experiments for 3.9 nm nanocrystals showed that all the XRD peak positions including the (200) peak recovered with pressure cycling to within 0.2% of its original value (see Fig. S6). The complete recovery of the (200) peak position indicated that the MT structure of the 3.9 nm nanocrystal was preserved with pressure cycling. In addition, the XRD peak widths for 3.9 nm nanocrystals showed the opposite trend as for 6 nm nanocrystals. The 3.9 nm XRD peak widths for (200) and (220) peaks increased by 16% and 23%, respectively, and remained at higher values after unloading. The peak width for (111) plane remained at about 2% of its initial value with pressure cycling. This indicated the introduction of surface nucleated partial dislocations (stacking faults) with pressure cycling.

The size-dependent MT to SC structural transformation can be analyzed in terms of the thermodynamic stability of the two structures. Howie and Marks represented the energy of a nanocrystal as:[36]

$$U = W_s + W_\gamma + W_{el} + H(V) \quad (1)$$

Where $W_s$, $W_\gamma$, $W_{el}$ and $H(V)$ are the energy due to surface stress, energy due to strain in the surface, elastic strain energy due to applied external pressure and nanocrystal geometry, and cohesive energy, respectively. Using this approach, it is found that the MT structure is stable at smaller sizes, the SC structure is stable at larger sizes and that the MT structure transforms into SC structure at a critical nanocrystal size of 7.2 nm at ambient pressure. At high pressure, the elastic strain energy and energy due to strain in the surface is modified to include additional energy input from the external pressure (see supplementary information). The transition size reduces with increasing pressure (see Fig. S7) and is 5.4 nm at 7.7 GPa (the maximum applied pressure in the experiments). This shows that it is thermodynamically favorable for 6 nm nanocrystals to be SC at high pressure, while it is favorable for 3.9 nm nanocrystals to be MT.

Similarly, MD simulations have shown that the MT structure is stable at smaller sizes and the SC structure is stable at larger sizes[9,13,14,37]. The MT structure transforms into the SC structure at a critical nanocrystal size of ~2-5 nm depending on the interatomic potential. This transition reflects the lower surface energy and higher lattice strain of MT structures. At high pressures, the MT structure is unfavorable compared to the SC structure due to its lower atomic packing fraction[18]. In our experiments, we went to a maximum pressure of ~7.7 GPa, this introduced ~33 meV per atom or 272 eV per nanocrystal additional energy into the system. The additional energy makes the MT to SC transition thermodynamically favorable at 6 nm size range where the energy gap between MT and SC structures is small.

Next, we consider the atomistic mechanism of the MT to SC transition at high pressure. Transformations in nanocrystals can occur through surface diffusion mediated mechanisms at elevated temperatures[20,21].

Diffusion is suppressed at high pressure and cannot be the mechanism for the MT to SC transformation in the nanocrystals[23]. At high pressure, the transformation can occur through a nondiffusive Mackay transformation or a dislocation/disclination mediated detwinning process. The Mackay transformation is a collective displacive atomic motion driven MT icosahedron to SC cuboctahedron transformation[18] (see Fig. S8). The Mackay transformation is highly symmetric and therefore requires low activation energy[38–40]. Simulation studies predict the dynamics of transformation using total energy calculation along the Mackay path[38,41,42] or MD simulations for small nanocrystals[43–45]. The MT to SC structural transformation can also proceed through dislocation or disclination mediated detwinning. Dislocation mediated detwinning was previously observed in large Pt nanocrystal under oxidative heating[16]. The SC grain nucleated at the surface of the nanocrystal and then grew when dislocation motion led to the retraction of twin boundaries. This transformation has also been observed to occur through the motion of disclinations[17].

The MT to SC transition is driven by deviatoric stresses caused by the nonhydrostatic pressure medium. The stress in the nanocrystals is higher along the loading axis (and the direction of imaging) than in the transverse direction. The difference between axial and transverse stress is termed differential stress. Differential stress in the sample chamber can be estimated using the lattice strain theory for FCC metals[46]. The maximum differential stress in 6 nm nanocrystals was ~2 GPa (see Fig. S9). We have previously shown that 3.9 nm nanocrystals can sustain dislocation activity due to the deviatoric stresses, while sustaining its twin boundary structures[26]. In order to understand the size-dependent stability of twin boundary structures, we performed MD simulations of 3.9 nm and 6 nm icosahedral nanocrystals (Fig. 5). Although the direct observation of structural transformation was not accessible in MD simulation due to the limited timescale, we were able to quantify the size-dependent pre-stress and to discover different twin boundary stabilities in small and large nanocrystals. While the angle between two non-parallel {111} surfaces is 70.53° in bulk FCC crystals, the twin boundaries in icosahedral nanocrystals form a 72° angle due to the five-fold symmetry, which inevitably induces pre-stress from the mismatch strain. The mismatch strain and resulting pre-stress inside icosahedral and decahedral MT nanocrystals can be approximated by the superposition of multiple finite-length disclinations. By assuming elastic isotropy and spherical surface, the pre-stress distribution inside MT icosahedral nanocrystal can be approximated as follows (see supplementary information).

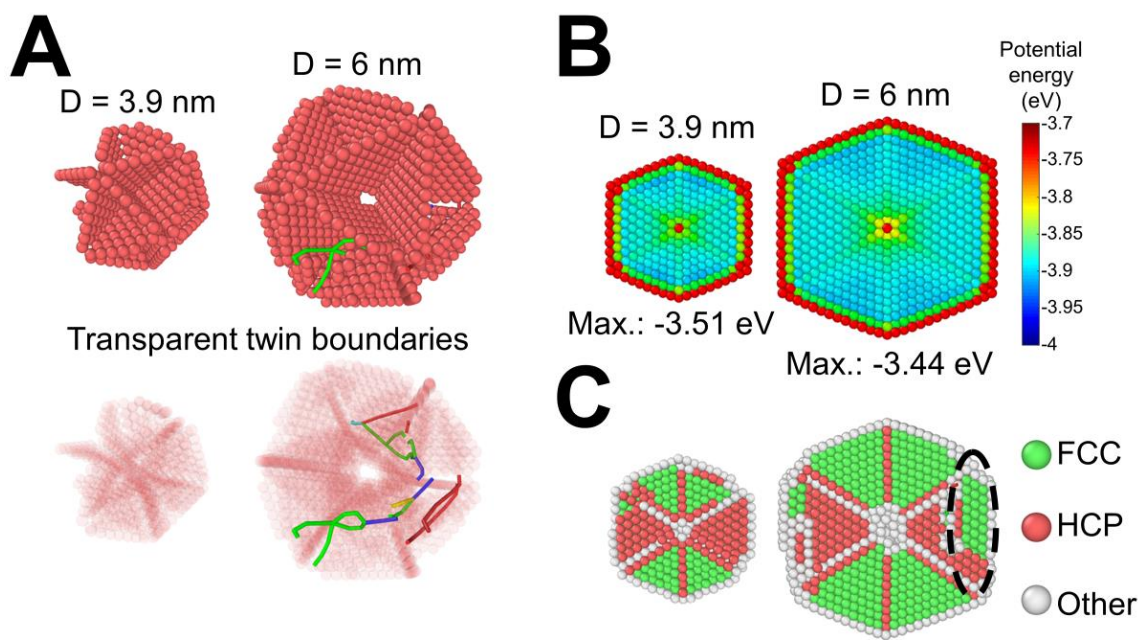

**Fig. 5. Atomistic simulation results of 3.9 nm and 6 nm icosahedral nanocrystals.** A) Twin boundary and dislocation structures in icosahedral nanocrystals using high temperature MD simulations. Dislocations are formed only in the 6 nm nanocrystal due to higher pre-stress. (green lines: Shockley partial dislocation, blue lines: full dislocation, red lines: dislocation blocked by twin boundaries). The red atoms are at twin boundaries. Atoms in regular FCC crystal positions are removed for visualization purposes. B) The atomic potential energy of pristine icosahedral nanocrystals. The 6 nm nanocrystal shows higher maximum potential energy (equivalently, higher pre-stress). C) Crystal structures of the nanocrystals after high temperature MD simulations. The twin boundary structure in 3.9 nm is preserved without noticeable distortion, while the twin boundary structure in 6 nm undergoes significant distortion.

$$\sigma_{rr} = \frac{4\mu\epsilon_I}{3}\left(\frac{1+\nu}{1-\nu}\right)\ln\left(\frac{r}{R}\right) - P \qquad (2)$$

where $\epsilon_I = 0.0615$, $\mu$ is the shear modulus, $\nu$ is the Poisson's ratio, R is the radius of the nanocrystal, P is the external pressure, and $r, \theta$ and $\phi$ are the spherical coordinates. The solution indicates pure compressive stress along the radial direction. The maximum value of compressive stress is found to be higher in the larger nanocrystal. Smaller nanocrystals are subjected to higher average strain energy and larger hydrostatic compression due to higher Laplace pressure from surface stress[36]. This is consistent with our ambient pressure XRD measurement where 3.9 nm shows a larger shift in the (111) peak position. Even though the theoretical analysis omits elastic anisotropy, the analytical solution with $\ln\frac{r}{R}$ dependence matches qualitatively well with the atomic potential energy distribution depicted in Fig. 5 B, which shows that 3.9 and 6 nm nanocrystals have higher strain energy density near the core and 6 nm nanocrystal has larger maximum atomic potential energy (i.e. higher pre-stress). Defect nucleation from the pristine twin structure is likely to initiate from the region of high pre-stress, so it is expected that defect nucleation occurs preferentially near the core of the MT nanocrystal. The MT structure in the larger nanocrystal is more susceptible to defect nucleation near the core because of its higher maximum pre-stress. The twin boundary structures with five-fold symmetry become progressively unstable for larger MT nanocrystals. We found that, even in the absence of any external stimuli, dislocation nucleation and distortion of twin boundaries were observed in 6 nm icosahedral nanocrystal in vacuum under relatively long high temperature MD simulation, while neither dislocation activity nor distortion of twin boundary structure is observed in the 3.9 nm nanocrystal due to smaller pre-stress (Fig. 5 C). These unstable twin boundary structures allow deviatoric stress on the 6 nm MT nanocrystal to drive the nondiffusive Mackay transformation or dislocation/disclination mediated detwinning.

In summary, we have used high-pressure XRD and post-compression TEM to provide the first evidence of deviatoric stress induced MT to SC structural transformation in nanocrystals. Energy calculations showed that the 6 nm MT nanocrystals become unstable at high pressures and the critical size for transition between MT and SC nanocrystals reduce with increasing pressure. MD simulations showed that the 6 nm MT nanocrystal was more susceptible to dislocation nucleation and had unstable twin boundaries. Kinetics of the process is governed by two possible nondiffusive paths - Mackay transformation and dislocation/disclination mediated detwinning. Deviatoric stresses in the nanocrystal drive the nondiffusive structural transformation. High-pressure SC nanocrystals were recovered after unloading, however, the nanocrystals quickly reverted back to MT state after redispersion in toluene solvent. The dynamics of recovery was captured using TEM which showed that the recovery was governed by surface recrystallization and rapid twin boundary motion. This study advances the understanding of stress driven structural transformation in nanoscale materials.

### ❖ Associated Content

**Supporting Information**.

- Detailed methods and experimental conditions with additional figures detailing data analysis, nanocrystal size distribution, simulated XRD patterns, TEM images, calculations for deviatoric stress and bulk modulus, derivation of thermodynamic MT to SC transition under pressure (PDF)
- TEM heating movie showing the nanocrystal twin boundary motion (MP4)

### ❖ Author Information

**Corresponding Author**
*Corresponding author:
  X. Wendy Gu
  452 Escondido Mall, Room 227,
  Stanford University, Stanford CA 94305
  650-497-3189
  xwgu@stanford.edu
**Author Contributions**
X.W.G. and A.P. conceived the idea and X.W.G. supervised the research of this work. A.P. synthesized the nanocrystals and M.T.K performed the TEM characterization. A.P., M.T.K., D.D., M.K. and A.D. performed the high-pressure XRD. A.P. performed the XRD simulation and analysis. S.L. and S.R. performed the MD simulations and analysis. A.P., S.L., S.R. and X.W.G. wrote the manuscript. All authors have given approval to the final version of the manuscript.

### ❖ Acknowledgement

X.W.G. and A.P. acknowledge financial support from Stanford start-up funds. The Advanced Light Source is supported by the Director, Office of Science, Office of Basic Energy Sciences, of the U.S. Department of Energy under Contract No. DE-AC02-05CH11231. Beamline 12.2.2 is partially supported by COMPRES, the Consortium for Materials Properties Research in Earth Sciences under NSF Cooperative Agreement EAR 1606856. Part of this work was performed at the Stanford Nano Shared Facilities (SNSF), supported by the National Science Foundation under award ECCS-1542152. M.T.K. is supported by the National Defense and Science Engineering Graduate Fellowship. D.D. is supported by the NSF Graduate Fellowship. S.L. and S.R. are supported by the Creative Materials Discovery Program (2016M3D1A1900038) through the National Research Foundation of Korea (NRF) funded by the Ministry of Science and ICT.

### ❖ Abbreviations

XRD, X-ray Diffraction; DAC, diamond anvil cell; MD, molecular dynamics; TEM, transmission electron microscopy; SC, single crystalline; MT, multiply twinned.

### ❖ References

(1) Xia, Y.; Xiong, Y.; Lim, B.; Skrabalak, S. E. Shape Controlled Synthesis of Metal Nanocrystals: Simple Chemistry Meets Complex Physics? *Angewandte*

Supplementary Material for

**Stress Induced Structural Transformations in Au Nanocrystals**

Abhinav Parakh[1], Sangryun Lee[2], Mehrdad T. Kiani[1], David Doan[3], Martin Kunz[4], Andrew Doran[4], Seunghwa Ryu[2] and X. Wendy Gu[3]*

[1]Materials Science and Engineering, Stanford University, Stanford, CA 94305, USA.

[2]Mechanical Engineering, KAIST, Yuseong-gu, Daejeon 34141, Republic of Korea.

[3]Mechanical Engineering, Stanford University, Stanford, CA 94305, USA.

[4]Advanced Light Source, Lawrence Berkeley National Lab, Berkeley 94720, USA.

**This PDF file includes:**

    Methods

    Analytical derivations

    Figs. S1 to S9

    Movie S1

    References

**Materials and Methods**

**Nanocrystal synthesis and characterization**

6 nm Au nanocrystals were synthesized according to Peng et al.[1]. 20 ml tetralin was combined with 24.3 ml 70% oleylamine and 200 mg HAuCl$_4$ in air at 25°C. A reducing solution of 1 mmol tert-butylamine-borane complex, 2 ml tetralin and 2.4 ml 70% oleylamine was then rapidly injected into the solution. The reaction was allowed to proceed for 1 hour. 100 μl dodecanethiol was added to the nanocrystals solution and then heated to 60° C for 15 min under N$_2$ gas. Nanocrystals were precipitated and washed using ethanol and redispersed in toluene. Nanocrystals were imaged using a FEI Tecnai G2 TEM at 200 keV accelerating voltage. Nanocrystal size distribution was determined from TEM images using ImageJ (Fig. S1).

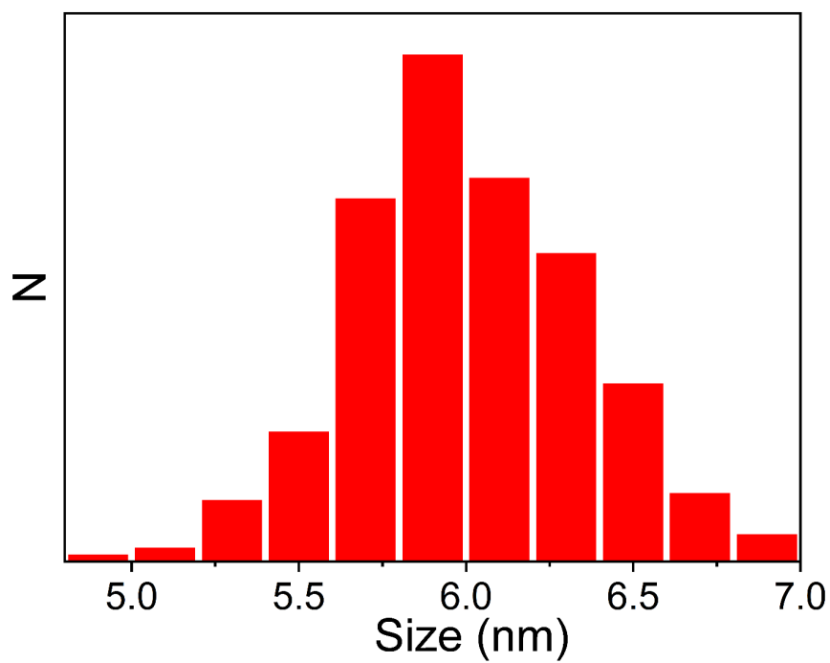

**Fig. S1. TEM size distribution of as-synthesized nanocrystals.** The average diameter of as-synthesized nanocrystals was 6.0±0.3 nm (300 particles measured).

**High pressure XRD**

Pressure-dependent measurements were performed in a Diacell© One20 DAC from Almax easyLab with ruby powder as a pressure calibrant. The diamonds had 500 µm culets and a T-301 stainless steel gasket with a 300 µm hole was used. Nanocrystals were drop casted on a glass slide to form a thick layer of gold nanocrystals. A small piece of the dried sample was loaded into the sample chamber with ruby powder and then the sample chamber was flooded with toluene. Toluene freezes at approximately 1.9 GPa and acts as a non-hydrostatic pressure medium[2]. The mean pressure was calculated from the shift in the R1 line[3].

XRD measurements were performed at beamline 12.2.2 at the Advanced Light Source at Lawrence Berkeley National Laboratory. The wavelength of the incident x-ray beam was fixed at 0.4976 Å and an x-ray spot size of 15 µm was used. Diffraction patterns were collected for 120 s using the Mar345 image plate detector. The sample to detector distance was calibrated using a $CeO_2$ standard. The 2D images were integrated to 1D plots using FIT2D software[4–6]. The XRD peak parameters were calculated by fitting the peaks to a combination of Gaussian and Lorentzian peak functions along with a high order polynomial for the background.

In addition to fitting errors, the contribution from instrumental broadening is important to consider. Even though the instrumental broadening is a fixed contribution, it is a lower limit on the accuracy of changes observed. The pixel resolution on the detector is 0.015º 2θ. The beamline reports a divergence of 0.5 mrad, which is equal to 0.028º 2θ broadening. The measured instrumental broadening (including the contribution from the sample placement) is 0.069º 2θ from the XRD standard at the beamline.

**MD simulations**

We employed LAMMPS software to run MD simulations of 3.9 nm and 6 nm icosahedral Au nanocrystals to investigate the size-dependent stability of twin boundary structures[7]. The atomistic interaction between the gold atoms was described using the EAM potential developed by H. Sheng[8]. First, we computed atomic potential energy distribution in two pristine nanocrystals to quantitatively assess geometrically-induced pre-stress and strain energy. Then, to test the stability of twin boundary structures, we conducted 900K simulations of two nanocrystals in vacuum under NVT ensemble for 500 ns with a time step of 1 fs. We did high temperature MD simulations to accelerate the defect activation by thermal fluctuation[9] because no defect activity was observed for both nanocrystals with 300K simulation due to limited timespan. For a clearer visualization of the dislocation and twin boundary structures, we relaxed the Au nanocrystal structure at 0K to remove the atomic displacement from thermal fluctuation. We used the open visualization tool (OVITO) to visualize the atomic configurations, and employed the dislocation extraction algorithm (DXA) to identify dislocations and stacking faults[10].

**Supplementary Text**

**Ambient pressure XRD**

Ambient pressure XRD for Au nanocrystals and bulk Au (from ICDD PDF: 00-004-0784) is plotted in Fig. S2.

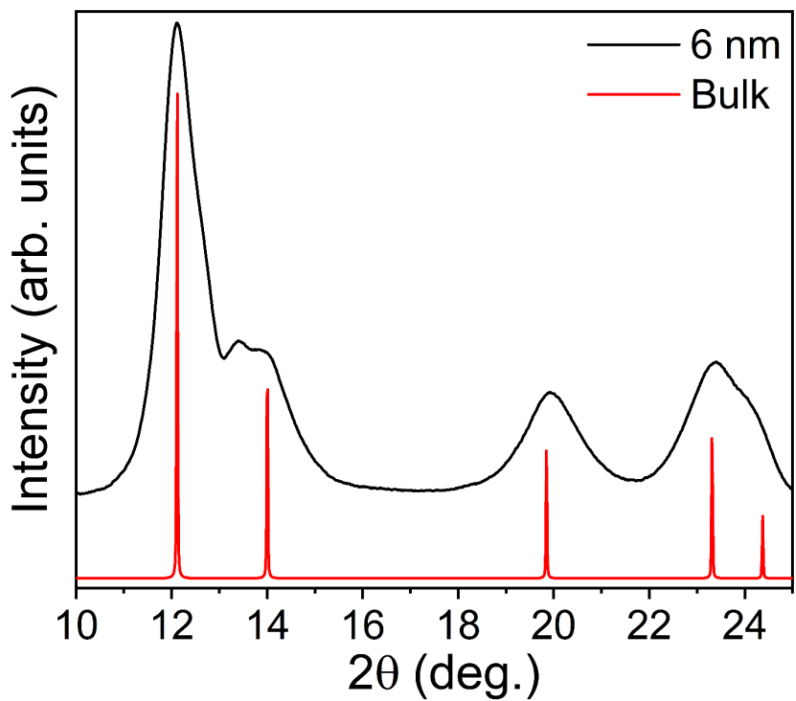

**Fig. S2. XRD patterns of 6 nm Au nanocrystal at ambient pressure and generated bulk Au (ICDD PDF:00-004-0784).**

**Bulk modulus calculation**

The unit cell volume was obtained at different pressures by fitting the (111), (220) and (311) diffraction peaks. The modulus that corresponds to the change in volume versus pressure was found to be 289 GPa (Fig. S3). This is significantly higher than the bulk modulus for bulk Au (~170 GPa), and the previously reported bulk modulus for Au nanocrystals with sizes from 10 to 20 nm (~196 GPa) [11,12]. The high value of the calculated modulus confirms the non-hydrostatic stress state within the diamond anvil cell and may have contributions from elastic size effects.

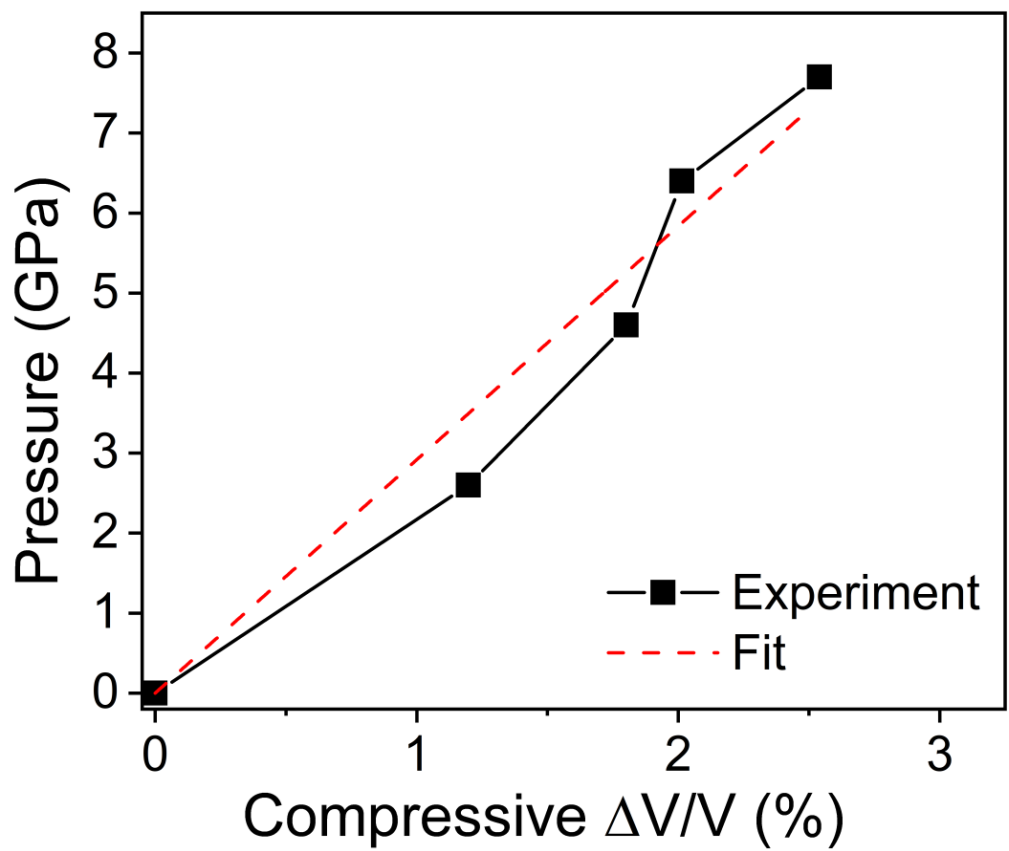

**Fig. S3. Changes in volumetric strain with pressure upon loading.** Linear fit was used to determine the bulk modulus for the Au nanocrystals.

**Debye scattering equation**

We constructed icosahedral MT and cuboctahedral SC Au nanocrystals of the sizes 5.8, 6 and 6.2 nm. The nanocrystals were equilibrated at 300K using MD simulations followed by 0 K energy minimization. The atomic positions were used to obtain powder XRD patterns by using the Debye equation. Two compositions were simulated: 80% MT and 20% SC, and 40% MT and 60% SC, to match the experimental pre and post compression distribution (Fig. S4). (200) peak was split into left and right peaks as in the experimental dataset and the ratio of left to right peak intensity is proportional to the fraction of MT nanocrystal in the sample.

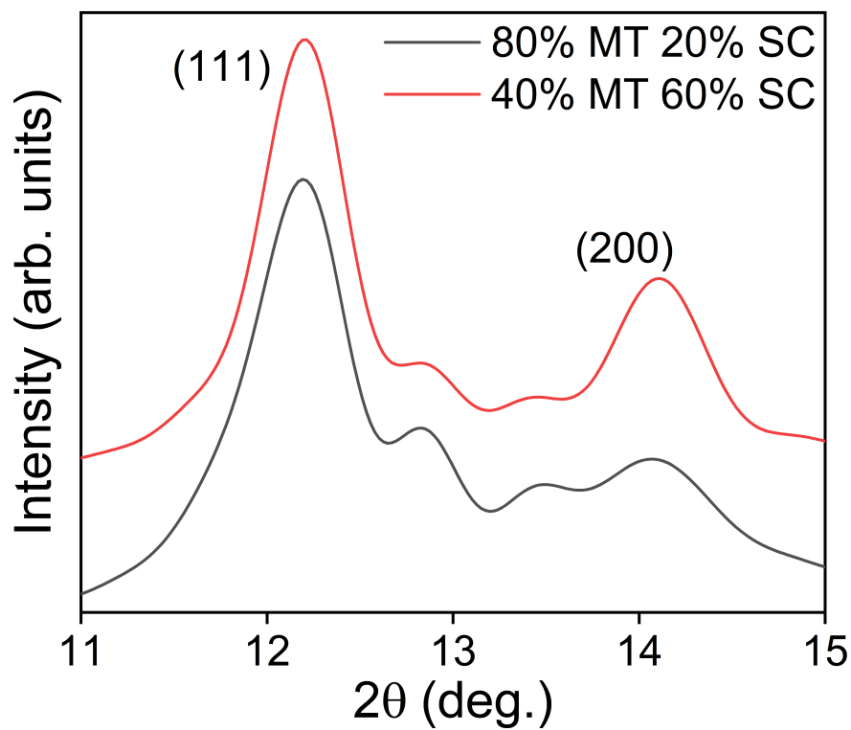

**Fig. S4. Simulated XRD patterns using Debye scattering equation for mixture of 6 nm icosahedral MT and cuboctahedral SC nanocrystals.** 80% MT and 20% SC showed higher ratio of left (200) to right (200) peak, and 40% MT and 60% SC showed lower ratio of left (200) to right (200) peak. In addition, the peak width for low MT mixture is smaller than for high MT mixture.

**Post-compression TEM images**

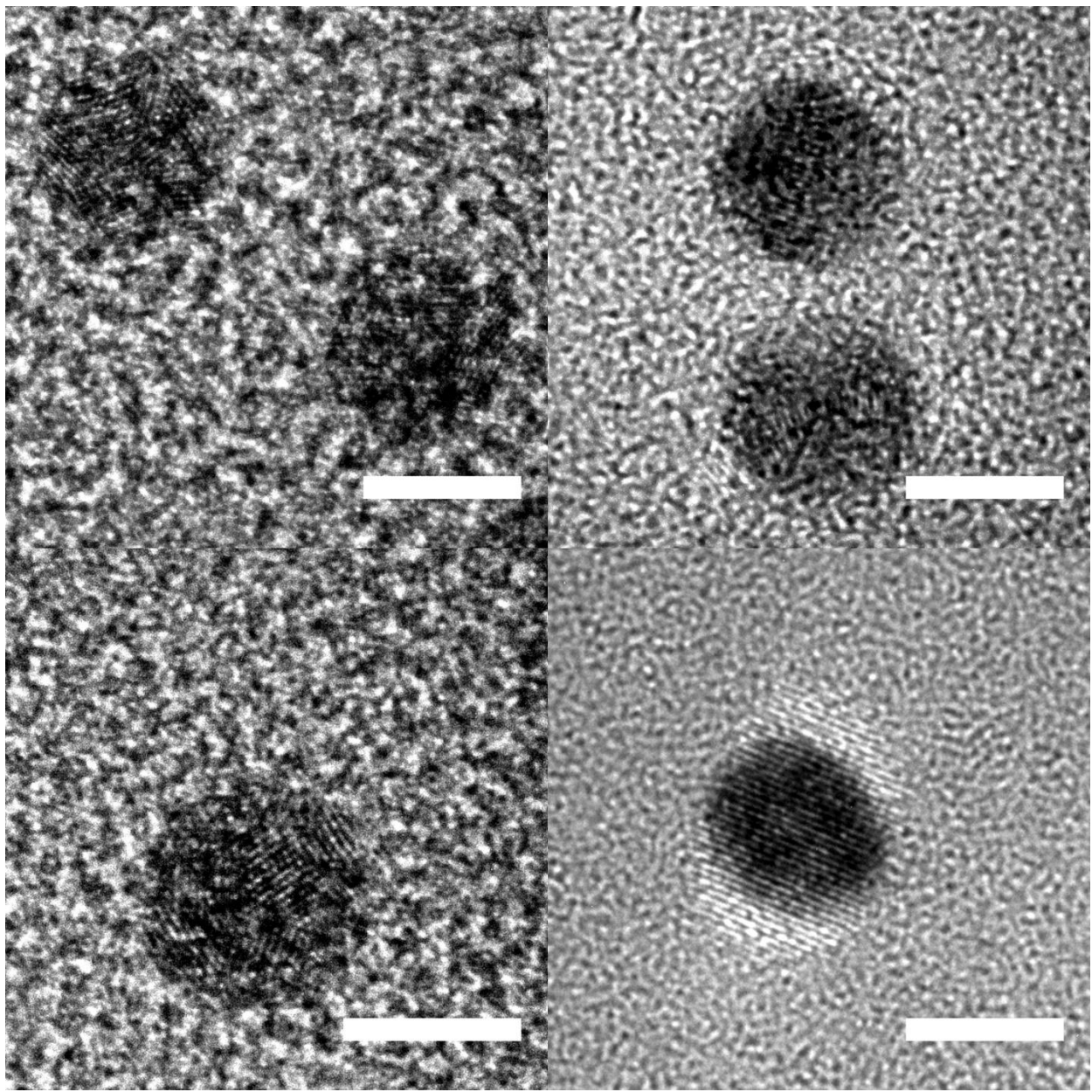

**Fig. S5. Post-compression TEM images of 6 nm Au nanocrystals after adding toluene to the grid after imaging.** Most SC nanocrystals reverted back to MT structure. Scale bar is 5 nm.

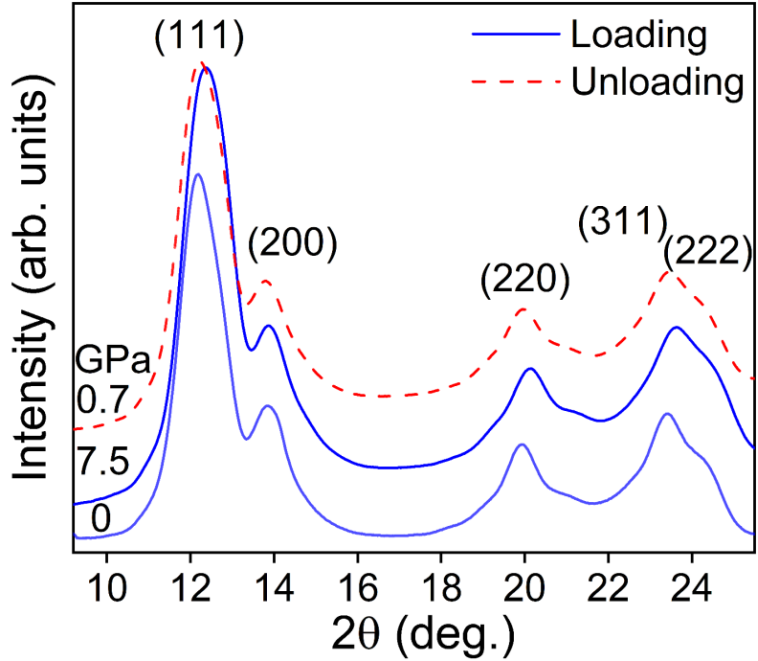

**Fig. S6. Experimental high-pressure XRD patterns for 3.9 nm nanocrystals[18].**

**Energy calculation for MT to SC transition**

The following notations and values are used in this calculation which are performed in spherical coordinates ($r, \theta$ and $\phi$).

Strains: $e_{rr}$, $e_{\theta\theta}$ and $e_{\phi\phi}$

Stresses: $\sigma_{rr}$, $\sigma_{\theta\theta}$ and $\sigma_{\phi\phi}$

Radial displacement: $u_r$

External pressure: $P$

Strain due to icosahedral geometry: $\epsilon_I$ (0.0615)

Poisson's ratio: $\nu$ (0.42)

Shear modulus: $\mu$ (27.4 GPa at ambient pressure and 34.6 GPa at 7.7 GPa)

Volume: $V$

Radius of the particle: $R$

Diameter of the particle: D

Elastic strain energy: $W_{el}$

IC denotes icosahedral

SC denotes single crystalline

Energy of straining the surface: $W_\gamma$

Surface energies: $\gamma_{111}$ (1.3 J/m² [14]), $\gamma_{100}$ (1.6 J/m² [14])

Twin boundary energy: $\gamma_t$ (0.04 J/m² [15,16])

Total strain in the surface: $\bar{e}_s$

Dimensionless parameter defined in ref. 17: $\epsilon_w$

Energy due to surface stresses: $W_s$

Total energy of the nanocrystal: $U$

Cohesive energy: H(V)

Here, we calculate the energy of a nanocrystal, and account for applied pressure:

$$U = W_s + W_\gamma + W_{el} + H(V) \qquad (2)$$

First, we calculate $W_\gamma$ and $W_s$. Following the derivation by Howie and Marks[13]:

$$e_{rr} = \frac{\partial u_r}{\partial r} \qquad (2)$$

$$e_{\theta\theta} = e_{\phi\phi} = \frac{u_r}{r} + \epsilon_I \qquad (3)$$

Equation for equilibrium:

$$\frac{\partial \sigma_{rr}}{\partial r} + \frac{2\sigma_{rr} - \sigma_{\theta\theta} - \sigma_{\phi\phi}}{r} = 0 \qquad (4)$$

Solving this with the boundary conditions that $u_r|_{r=0} = 0$ and $\sigma_{rr}|_{r=R} = -P$:

$$u_r = \frac{2\epsilon_I}{3}\left(\frac{1-2\nu}{1-\nu}\right) r \ln(r) + \left(-\frac{P(1-2\nu)}{2\mu(1+\nu)} - \frac{2}{3}\epsilon_I - \frac{2}{3}\epsilon_I \left(\frac{1-2\nu}{1-\nu}\right) \ln(R)\right) r \qquad (5)$$

Using this we get,

$$e_{rr} = \frac{2}{3}\epsilon_I \left(\frac{1-2\nu}{1-\nu}\right)\left[\ln\left(\frac{r}{R}\right) + 1\right] - \frac{2}{3}\epsilon_I - \frac{P(1-2\nu)}{2\mu(1+\nu)} \qquad (6)$$

$$e_{\theta\theta} = e_{\phi\phi} = \frac{2}{3}\epsilon_I \left(\frac{1-2\nu}{1-\nu}\right) \ln\left(\frac{r}{R}\right) + \frac{1}{3}\epsilon_I - \frac{P(1-2\nu)}{2\mu(1+\nu)} \qquad (7)$$

$$\sigma_{rr} = -P + \frac{4}{3}\epsilon_I \mu \left(\frac{1+\nu}{1-\nu}\right) \ln\left(\frac{r}{R}\right) \qquad (8)$$

$$\sigma_{\theta\theta} = \sigma_{\phi\phi} = \sigma_{rr} + \frac{2\mu\epsilon_I}{3}\left(\frac{1+\nu}{1-\nu}\right) \qquad (9)$$

Elastic strain energy:

$$W_{el} = \frac{1}{2}\left(\sigma_{rr}e_{rr} + \sigma_{\theta\theta}e_{\theta\theta} + \sigma_{\phi\phi}e_{\phi\phi}\right) \qquad (10)$$

For IC nanocrystal:

$$W_{el}^{IC} = \left(\frac{(1-2\nu)3P^2}{(1+\nu)4\mu} + \frac{2\epsilon_I^2 \mu(\nu+1)}{3(1-\nu)}\right) V \qquad (11)$$

Where,

$$V = \frac{4\pi}{3} R^3 \qquad (12)$$

Similarly, for SC nanocrystal:

$$W_{el}^{SC} = \left(\frac{(1-2v)3P^2}{(1+v)4\mu}\right)V \tag{13}$$

Energy due to surface strain:

$$W_\gamma = \epsilon_w \gamma_{111} V^{\frac{2}{3}}(1+\bar{e}_s) \tag{14}$$

Where,

$$\epsilon_w^{IC} = \{67.5\sqrt{3}[(1+3\eta)^3 - 24\eta^3]\}^{\frac{1}{3}} \tag{15}$$

$$\epsilon_w^{SC} = [108\sqrt{3}(1-3\beta^3)]^{\frac{1}{3}} \tag{16}$$

$$\eta = \frac{\gamma_t}{2\gamma_{111}} \text{ and } \beta = 1 - \frac{\gamma_{100}}{\sqrt{3}\gamma_{111}} \tag{17}$$

$$\bar{e}_s^{IC} = \frac{2}{3}\epsilon_I - \frac{P(1-2v)}{\mu(1+v)} \tag{18}$$

$$\bar{e}_s^{SC} = -\frac{P(1-2v)}{\mu(1+v)} \tag{19}$$

Energy due to surface stresses:

$$W_s = -\frac{\epsilon_w^2 \gamma_{111}^2 V^{\frac{1}{3}}}{6\mu}\left(\frac{1-2v}{1+v}\right) \tag{20}$$

Total energy for nanocrystals is given by:

$$U = W_s + W_\gamma + W_{el} + H(V) \tag{21}$$

$$U = -\frac{\gamma_{111}^2 V^{\frac{1}{3}}}{6\mu}\left(\frac{1-2v}{1+v}\right)(\epsilon_w^2) + \gamma_{111} V^{\frac{2}{3}}[\epsilon_w(1+\bar{e}_s)] + W_{el} + H(V) \tag{22}$$

Energy difference between SC and IC nanocrystals:

$$\Delta U = -\frac{\gamma_{111}^2 \left(\frac{4\pi}{3}R^3\right)^{\frac{1}{3}}}{6\mu}\left(\frac{1-2\nu}{1+\nu}\right)\Delta(\epsilon_w^2) + \gamma_{111}\left(\frac{4\pi}{3}R^3\right)^{\frac{2}{3}}\Delta[\epsilon_w(1+\bar{e}_s)] + \Delta W_{el} \quad (23)$$

Setting $\Delta U = 0$ gives the equilibrium $R$ and D for SC to IC transition. At ambient pressure ($P = 0.1$ MPa) we get D = 7.2 nm and at high pressure ($P = 7.7$ GPa) we get D = 5.4 nm (Fig. S7).

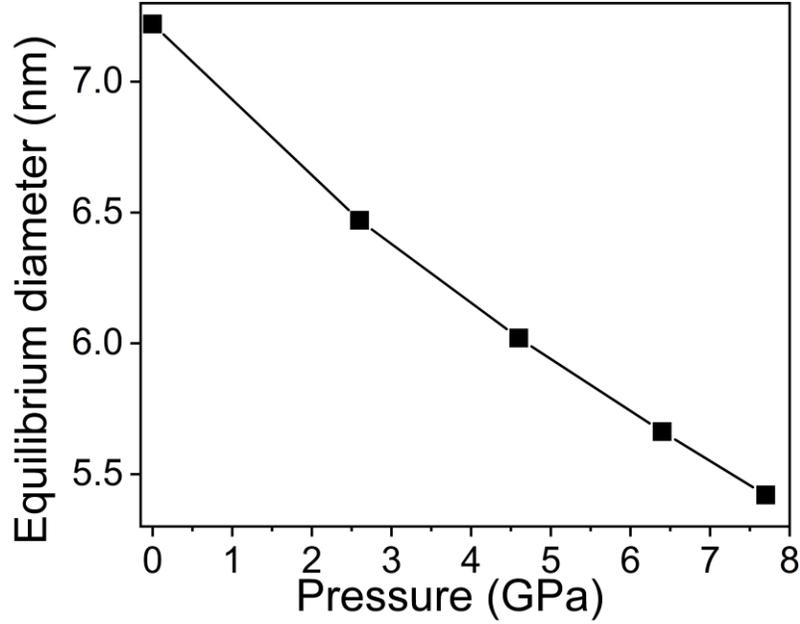

**Fig. S7. Pressure dependence of equilibrium diameter for SC to IC transition, plotted using eqn. (23).**

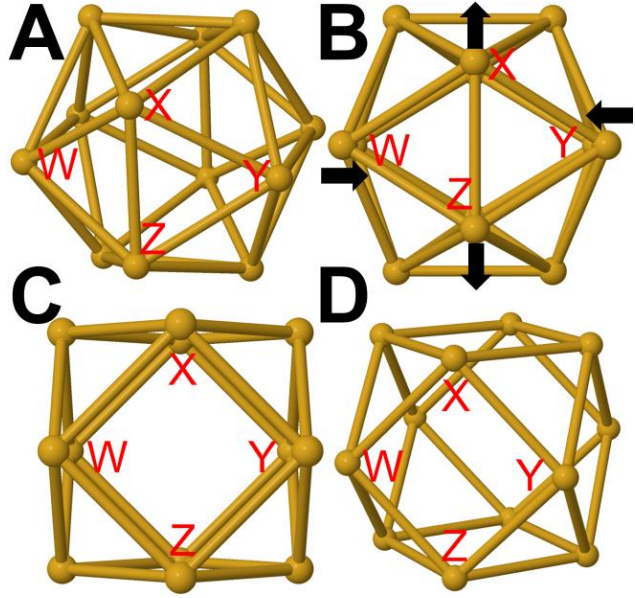

**Fig. S8. Continuous displacive Mackay transformation from icosahedral to cuboctahedral.** A) Initial multiply twinned icosahedral shape. B) Direction of motion of atoms and the applied force by non-hydrostatic pressure. C) Bond between X and Z atoms is broken, and all the atoms are relaxed. D) Final transformed cuboctahedral shape.

**Calculation of deviatoric stress**

The difference between the axial and radial stress (t) is calculated using lattice strain theory at each pressure [19]. First, the quantity $Q(hkl)$ is calculated for the (111), (220), and (311) peaks:

$$Q(hkl) = \frac{[a_m(hkl) - a_p]}{a_p(1 - 3\sin^2\theta_{hkl})} \qquad (24)$$

Where $a_m(hkl)$ is the lattice parameter from the experimental data (non-hydrostatic pressure), $a_p$ is the expected lattice parameter of Au under hydrostatic pressure, and $\theta_{hkl}$ is the experimental XRD peak position. $a_p$ is calculated by using 196 GPa as the bulk modulus of Au nanocrystals [12], and an effective hydrostatic pressure that is the sum of the applied pressure (measured from Ruby peak shift) and the pressure due to surface stress which was determined by the ambient pressure peak shift compared to bulk Au (Fig. S2).

t is then calculated as:

$$t = (6G)\langle Q(hkl)\rangle f(x) \quad (35)$$

$\langle Q(hkl)\rangle$ is the average of Q(111), Q(220) and Q(311). G is the shear modulus at the hydrostatic pressure[20]. f(x) is equal to:

$$f(x) = \frac{A}{B} \quad (26)$$

Where A and B are constants that are defined as:

$$A = \frac{2x+3}{10} + \frac{5x}{2(3x+2)} \quad (27)$$

$$B = \alpha[x - 3(x-1)\langle \Gamma(hkl)\rangle] + \frac{5x(1-\alpha)}{3x+2} \quad (28)$$

$$x = \frac{2(S_{11} - S_{12})}{S_{44}} \quad (29)$$

$$\Gamma(hkl) = \frac{(h^2k^2 + k^2l^2 + l^2h^2)}{(h^2 + k^2 + l^2)^2} \quad (30)$$

α is equal to 0.5 (in between Reuss (iso-stress) and Voigt (iso-strain) conditions) [21]. $\Gamma(hkl)$ is calculated for the (111), (220) and (311) peaks and then averaged to find $\langle\Gamma(hkl)\rangle$. t as a function of applied pressure is shown in Fig. S9.

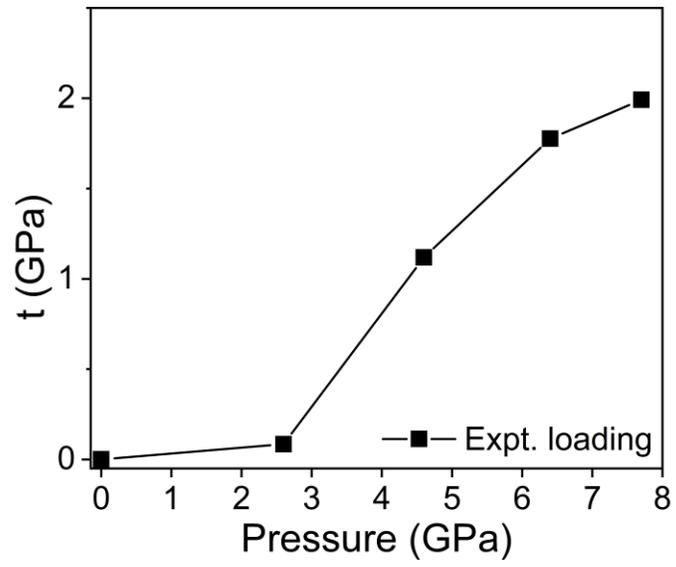

**Fig. S9. Pressure dependence of calculated deviatoric stress from experiments.**

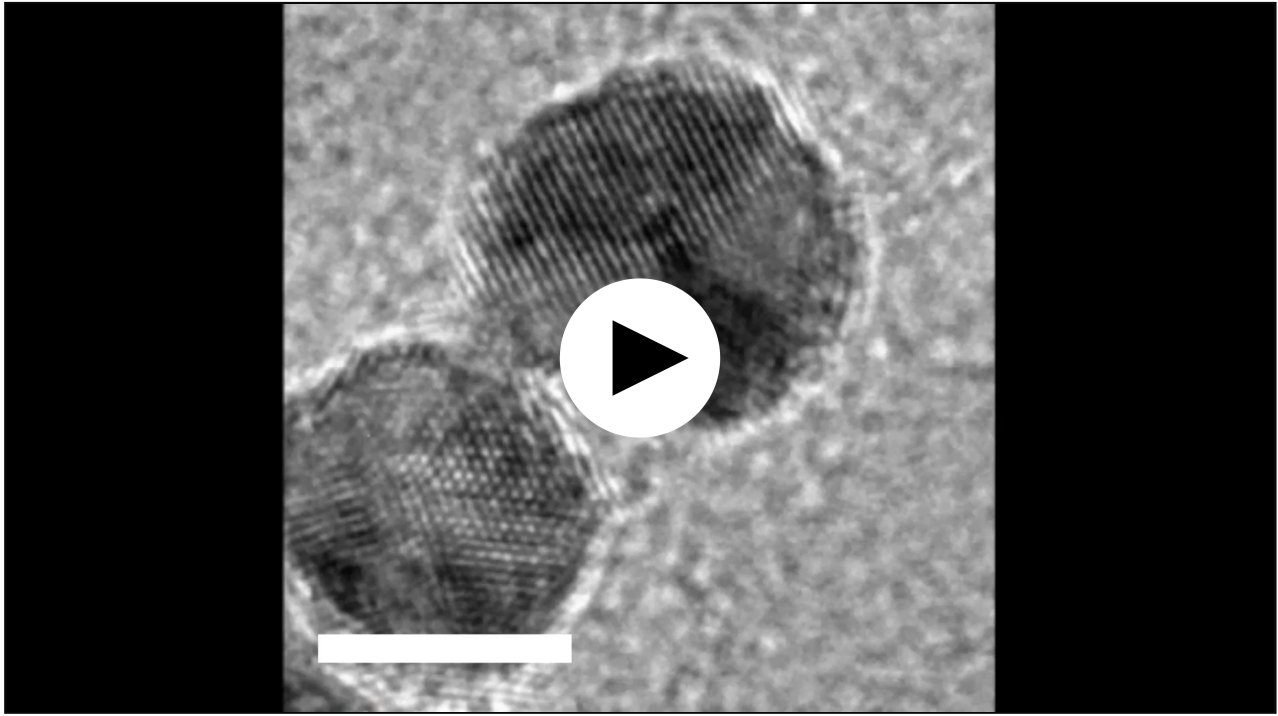

**Movie S1. In-situ TEM video of post-compression nanocrystals under high dose electron beam.** The heating caused rapid motion of twin boundary and sintering of the nanocrystals. Scale bar is 5 nm.